\documentclass[prb,twocolumn,superscriptaddress,float,aps]{revtex4-2}
\usepackage{bm,color,amsmath,amssymb,mathrsfs,latexsym,graphicx,psfrag,tabularx}

\usepackage[hidelinks,colorlinks,linkcolor=blue,
citecolor=blue,urlcolor=blue]{hyperref}

\newcommand{\beq}{\begin{equation}}
\newcommand{\eneq}{\end{equation}}

\def\kk{\mathbf{k}}

\usepackage{etoolbox}
\makeatletter
\patchcmd{\@maketitle}{\@author}{\@author\show\@thanks}{}{}
\makeatother

\usepackage{dsfont}

\begin{document}

\title{Nonlinear Tellegen Limit}

\author{Abhinava Chatterjee}
\email{abhinava.chatterjee@psu.edu}
\affiliation{%
   Department of Physics, The Pennsylvania State University, University Park, Pennsylvania 16802, USA
}
\author{Nikhil Kalyanapuram}
\email{nkalyanapuram@psu.edu}
\affiliation{%
   Applied Research Laboratory, The Pennsylvania State University, State College, Pennsylvania 16801, USA
}

\begin{abstract}
The Tellegen limit is the fundamental electromagnetic 
stability bound on magnetoelectric media. We show that 
nonlinear magnetoelectric coupling gives rise to a new 
Tellegen limit, which we term the \textit{nonlinear 
Tellegen limit}. 
Unlike the linear Tellegen limit, which is fixed by 
material parameters, the nonlinear Tellegen limit is 
field-tunable — a static electric or magnetic field drives the 
system toward electromagnetic instability at a material-specific critical field determined by the
symmetry-allowed nonlinear magnetoelectric couplings and electromagnetic response. The approach to this 
limit is accompanied by a field-tunable linear birefringence, whose 
maximum is bounded by the nonlinear Tellegen limit — beyond which the medium 
becomes electromagnetically unstable. We demonstrate 
the nonlinear Tellegen limit and the field-space 
stability phase diagram in two magnetically ordered 
material systems — a $d$-wave altermagnet and an 
M-type hexagonal ferrite — showing that the symmetry 
of the magnetic point group governs both the structure 
of the nonlinear magnetoelectric tensor and the 
resulting electromagnetic instability.
\end{abstract}

\date{\today}


\maketitle

\textit{Introduction}.
Magnetoelectric media — materials in which electric and 
magnetic degrees of freedom are coupled \cite{curie1894symetrie,landau2013electrodynamics,dzyaloshinskii1959magnetoelectrical,astrov1961magnetoelectric,astrov1960magnetoelectric,rado1961observation,o1962electrodynamics,o1963field,fuchs1965wave} — exhibit a 
remarkable range of electromagnetic phenomena, from 
nonreciprocity \cite{toyoda2019nonreciprocal,toyoda2021nonreciprocal,krowne1995nonreciprocity,nikitchenko2021magnetic,tokura2018nonreciprocal,arima2008magneto}, to novel optical responses \cite{schmid2000magnetoelectric,fiebig2005revival,srinivasan2010magnetoelectric,eremenko1987magneto,hu2019opportunities,prudencio2016optical} and 
multiferroic order \cite{cheong2018broken,mostovoy2024multiferroics,rivera2009short,pyatakov2012magnetoelectric,eerenstein2006multiferroic,fetisov2024nonlinear,scott2013room,schmid2008some,fiebig2005revival,
fiebig2016evolution,spaldin2019advances}. A fundamental 
constraint governing all such media is the Tellegen 
limit, where the magnetoelectric susceptibility $\alpha$ is 
bounded above by the geometric mean of the dielectric 
and magnetic responses, $\alpha^2 \leq \epsilon\mu$ \cite{tellegen1948gyrator,brown1968upper}, a 
condition whose saturation signals the onset of 
electromagnetic instability~\cite{seidov2025unbounded}. While extensively studied in the linear 
regime~\cite{fiebig2005revival}, the fate 
of this bound in the presence of nonlinear 
magnetoelectric coupling has remained unexplored. 

Nonlinear magnetoelectric effects, in which the 
polarization or magnetization responds quadratically 
or cubically to applied fields \cite{fiebig2005revival,de2025light,jia2026nonlinear,hu2025nonlinear}, have been observed 
in a growing class of magnetically ordered materials. 
In hexagonal ferrites, giant nonlinear magnetoelectric 
coupling has been measured in M-type barium and 
strontium hexaferrites~\cite{zavislyak2016current,
liu2021nonlinear,popov2022plane} and Y-type 
hexaferrites~\cite{popov2020nonlinear}, with 
coefficients an order of magnitude larger than those 
of any known linear magnetoelectric. Similar studies have demonstrated large nonlinear magnetoelectric effects in buckled-honeycomb antiferromagnets \cite{lee2020highly}. In the emerging class of altermagnets — collinear 
magnets with momentum-dependent spin splitting and 
broken time-reversal 
symmetry~\cite{smejkal2022beyond,smejkal2022emerging} 
— recent 
theoretical work has shown that altermagnets support a nonlinear 
magnetoelectric effect characterized by a second-order 
response to an applied electric 
field~\cite{oike2024nonlinear,bhowal2024ferroically,yang2025nonlinear,yang2026nonlinear,
smejkal2024altermagnetic}, making them natural 
candidates for the nonlinear Tellegen limit. The $d$-wave altermagnets, including RuO$_2$~\cite{berlijn2017itinerant,smejkal2022beyond,
fedchenko2024observation,feng2022anomalous} and 
MnF$_2$ \cite{yamani2010neutron,yuan2020giant,vsmejkal2020crystal,egorov2021colossal}
exemplify this symmetry structure. Although the altermagnetic character of RuO$_2$ remains 
experimentally contested~\cite{kessler2024absence,
smolyanyuk2024fragility}, MnF$_2$ provides an uncontested realization of the same $d$-wave magnetic point group.  Yet the consequences of nonlinear magnetoelectric coupling in these materials for electromagnetic stability have gone entirely unexamined. Recent experimental and theoretical advances in 
photonic metamaterials have demonstrated that the 
Tellegen limit can be approached through deliberate 
geometric engineering of artificial 
structures~\cite{seidov2025unbounded,yang2025gigantic,saltykova2025analytical,kamenetskii2009tellegen}. 
Here, we show that the same limit can be reached 
dynamically in natural magnetically ordered materials 
by applying a static field — without any artificial 
structuring. 

In this work, we show that nonlinear magnetoelectric 
coupling gives rise to a fundamentally new phenomenon: a 
\textit{nonlinear Tellegen limit}, which is tunable by an applied static electric or magnetic 
field. Specifically, a static electric field $\mathbf{E}^{(0)}$ 
induces an effective magnetoelectric tensor 
$\alpha_{ij} = \beta_{ikj}E_k^{(0)}$ and a static magnetic field $H^{(0)}$ induces a correction to the dielectric tensor $\epsilon_{ij}$ through the cubic 
coupling, driving the system toward electromagnetic 
instability at a material-specific critical field $E^*(H^*)$. We further demonstrate that the approach to this limit has a striking experimental 
consequence: the linear birefringence becomes electrically 
tunable, with its maximum bounded by the nonlinear Tellegen limit, beyond which the medium 
is electromagnetically unstable. We establish these phenomena in two material systems with distinct magnetic 
symmetries: the $d$-wave altermagnet MnF$_2$, and M-type 
hexagonal ferrite. In MnF$_2$, the nonlinear Tellegen limit 
is driven by an applied electric field and is accompanied 
by electrically tunable linear birefringence. In hexagonal ferrites, a richer phenomenology 
emerges: in-plane magnetic fields drive instability while 
out-of-plane magnetic fields stabilize the medium, a competition 
captured in a field-space phase diagram. Our results establish the nonlinear Tellegen limit as a 
new class of electromagnetic instability in magnetically ordered materials, with experimentally accessible signatures in optical 
measurements.

\begin{figure}
\centering
\includegraphics[width=\columnwidth]{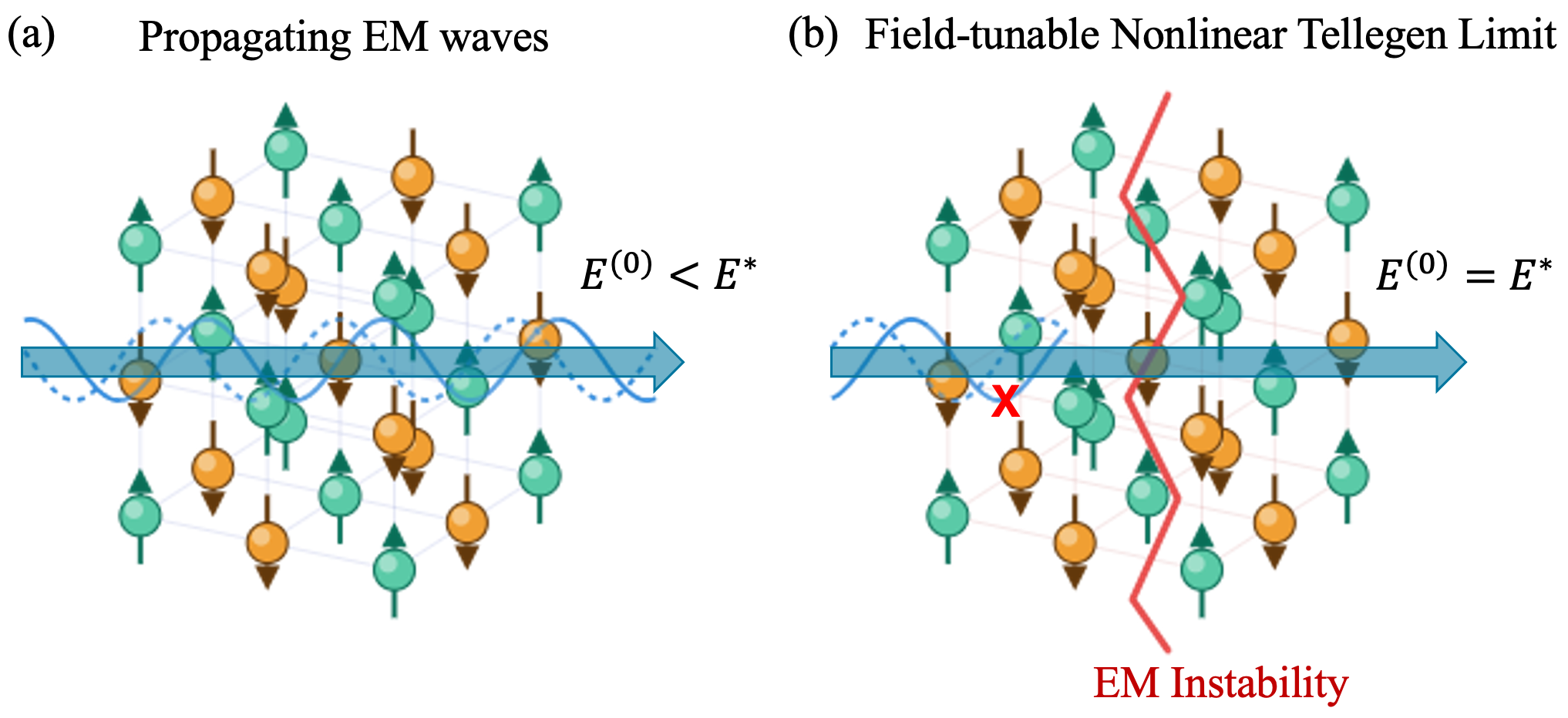}
\caption{ (a) Propagating electromagnetic waves in a centrosymmetric magnetic crystal when the static field $E^{(0)}$ is smaller than the nonlinear Tellegen limit $E^*$. (b) The electromagnetic instability when the static field $E^{(0)}$ reaches the nonlinear Tellegen limit $E^*$.  
}
\label{fig:Setup}
\end{figure}

\textit{Nonlinear Magnetoelectric Coupling}.
We consider centrosymmetric magnetically ordered materials 
whose free energy contains a leading cubic magnetoelectric 
coupling of the form
\begin{equation}\label{eq:Free0}
    F = \frac{1}{2}\epsilon_{ij}E_{i}E_{j} 
      + \frac{1}{2}\mu_{ij}H_{i}H_{j}  
      + \frac{1}{2}\beta_{ijk}H_{i}E_{j}E_{k},
\end{equation}
where $\epsilon_{ij} = \epsilon_{ji}$ and $\mu_{ij} = \mu_{ji}$ 
are the dielectric and magnetic permeability tensors, and 
$\beta_{ijk} = \beta_{ikj}$ are the nonlinear magnetoelectric 
coefficients, symmetric under exchange of the last two indices. 
We neglect terms of order $H^3$, in light of their expected relative smallness. The resulting constitutive relations are $P_i = \partial F/\partial E_i$ and 
$M_i = \partial F/\partial H_i$:
\begin{equation}
    \begin{aligned}
        P_{i} & = \epsilon_{ij}E_{j} + \beta_{jki}H_{j}E_{k}, \\
        M_{i} & = \mu_{ij}H_{j} + \tfrac{1}{2}\beta_{ijk}E_{j}E_{k}.
    \end{aligned}
\end{equation}
We expand the fields around a static background, 
$E_i = E^{(0)}_i + \tilde{E}_i\cos(\kk\cdot\mathbf{x} - \omega t)$ 
and likewise for $H_i$. In the limit 
$|\tilde{\mathbf{E}}| \ll |\mathbf{E}^{(0)}|$, the oscillatory 
response acquires a bianisotropic character:
\begin{align}
    P^{(1)}_i &= \left(\tilde{\epsilon}_{ij}\tilde{E}_{j} 
               + \alpha_{ji}\tilde{H}_{j}\right)
               \cos(\kk\cdot\mathbf{x} - \omega t), \\
    M^{(1)}_i &= \left(\mu_{ij}\tilde{H}_{j} 
               + \alpha_{ij}\tilde{E}_{j}\right)
               \cos(\kk\cdot\mathbf{x} - \omega t),
\end{align}
with field-renormalized susceptibilities
\begin{equation}
    \tilde{\epsilon}_{ij} = \epsilon_{ij} + \beta_{kji}H_k^{(0)},
    \qquad
    \alpha_{ij} = \beta_{ikj}E_k^{(0)}.
\end{equation}
The symmetry structure of these tensors is fixed by inversion; 
since $P_i$ is a polar vector and 
$\alpha_{ij}\tilde{H}_j \sim P^{(1)}_i$, 
the induced magnetoelectric tensor $\alpha_{ij}$ must be odd 
under inversion, which requires it to be linear in 
$\mathbf{E}^{(0)}$, consistent with 
$\alpha_{ij} = \beta_{ikj}E_k^{(0)}$. Further, corrections to 
$\tilde{\epsilon}_{ij}$ must be even under inversion and 
therefore involve only $\mathbf{H}^{(0)}$. Since $M_i$ is 
an axial vector and even under inversion, $\mu_{ij}$ cannot 
acquire corrections linear in $\mathbf{E}^{(0)}$; the 
magnetic permeability is therefore unrenormalized at this 
order. Finally, since $\beta_{ijk} = \beta_{ikj}$, 
$\tilde{\epsilon}_{ij}$ is symmetric, while $\alpha_{ij}$\footnote{This is in agreement with conventional expectations when calculating $\alpha_{ij}$ from a free energy expansion.} is 
generically asymmetric since $\beta_{ikj} \neq \beta_{jki}$. 

The nonlinear coupling $\beta_{ijk}$ thus generates an 
effective magnetoelectric tensor $\alpha_{ij}$ controlled 
entirely by the applied field, furnishing a nonlinear analog of the 
linear magnetoelectric effect, absent in the linear regime. The stability of the resulting bianisotropic medium is governed by the positive definiteness of the nonlinear Tellegen matrix
\begin{equation}\label{eq:tellegen}
    \mathcal{D} \equiv \tilde{\epsilon} - \alpha^T \mu^{-1}\alpha,
    \qquad 
    \lambda_{\min}(\mathcal{D})>0,
\end{equation}
where $\lambda_{\min}(\mathcal{D})$ denotes the lowest eigenvalue of
$\mathcal{D}$. The nonlinear Tellegen limit is reached when
$\lambda_{\min}(\mathcal{D})=0$, signaling the onset of electromagnetic
instability \cite{tellegen1948gyrator,brown1968upper}.

\textit{Nonlinear Tellegen Limit}.
We demonstrate the nonlinear Tellegen limit in two material 
systems: an insulating $d$-wave altermagnet such as 
MnF$_2$ \cite{yamani2010neutron,yuan2020giant,vsmejkal2020crystal,egorov2021colossal}, and 
M-type hexagonal ferrite \cite{zavislyak2016current,
liu2021nonlinear,popov2022plane}, a well-characterized 
nonlinear magnetoelectric medium. Both sets of materials forbid a 
linear magnetoelectric response by symmetry while permitting 
cubic magnetoelectric contributions to their free energies.

(i) \textit{$d$-wave altermagnet: MnF$_2$}. --- The magnetic point group of MnF$_2$ is $4^{\prime}/mm^\prime m$, with generators $\hat{C}_{4z}\hat{T}$, $\hat{m}_x$, 
$\hat{m}_y$, and spatial inversion $\hat{P}$. These symmetry 
constraints permit nonzero nonlinear magnetoelectric 
coefficients $\beta_{xxz} = \beta_{xzx} = -\beta_{yzy} = - 
\beta_{yyz} \equiv \beta$, with all couplings involving 
$H_z$ forbidden by symmetry. We consider only a static electric field,  i.e., 
$E_i^{(0)} \neq 0,  H_i^{(0)} =0$. By introducing dimensionless fields
$\mathcal{E}_{i} = \beta E_{i}^{(0)}/\sqrt{\epsilon_{xx}\mu_{xx}}$
with $\mathcal{E}_\parallel^2 = \mathcal{E}_x^2 + \mathcal{E}_y^2$,
we solve for the eigenvalues of $\mathcal{D}$. The full set of
eigenvalues of $\mathcal{D}$ in the presence of both
$E_i^{(0)},  H_i^{(0)}$ is given in Sec. S2 of the Supplementary
Material (SM) \cite{SM}. Fig.~\ref{fig:RuO}(a) plots the lowest eigenvalue
$\lambda_\text{min}(\mathcal{D})$ rescaled by $\epsilon_{xx}$,
given by $\widetilde{\lambda}_\text{min}
= \lambda_\text{min}/\epsilon_{xx}$, in the absence of an external
magnetic field $\textbf{H}^{(0)}=0$. The red curve shows
$\widetilde{\lambda}_\text{min}$ vs.\ $\mathcal{E}_\parallel$
at $\mathcal{E}_z=0$. The three rescaled eigenvalues in this limit are
$\{1,1,\epsilon_{zz}/\epsilon_{xx}-\mathcal{E}_\parallel^2\}$.
For MnF$_2$, $\epsilon_{xx}>\epsilon_{zz}$ \cite{seehra1986effect},
which implies that the lowest rescaled eigenvalue is given by
$\widetilde{\lambda}_\text{min}
=\epsilon_{zz}/\epsilon_{xx}-\mathcal{E}_\parallel^2$.
The nonlinear Tellegen limit is reached at
$\mathcal{E}_\parallel
=\pm\sqrt{\epsilon_{zz}/\epsilon_{xx}}$. The blue curve shows the same quantity
$\widetilde{\lambda}_\text{min}$ vs.\ $\mathcal{E}_z$ at
$\mathcal{E}_\parallel=0$. In this case,
$\widetilde{\lambda}
=\{\epsilon_{zz}/\epsilon_{xx},1-\mathcal{E}_z^2,
1-\mathcal{E}_z^2\}$ and
$\widetilde{\lambda}_\text{min}
=\text{min}\left(\epsilon_{zz}/\epsilon_{xx},
1-\mathcal{E}_z^2\right)$.
When
$|\mathcal{E}_z|
<\sqrt{1-\epsilon_{zz}/\epsilon_{xx}}$,
the lowest eigenvalue is a constant, as seen by the horizontal blue line
in Fig.~\ref{fig:RuO}(a). Increasing the magnitude of the out-of-plane
electric field beyond
$\sqrt{1-\epsilon_{zz}/\epsilon_{xx}}$
leads to a field-dependent behavior of
$\widetilde{\lambda}_\text{min}$, and the instability occurs at
$\mathcal{E}_z=\pm1$. Two features are evident: 
the nonlinear Tellegen limit is electrically tunable, and 
its critical value depends sensitively on the direction of 
the applied field.

\begin{figure}
\centering
\includegraphics[width=\columnwidth]{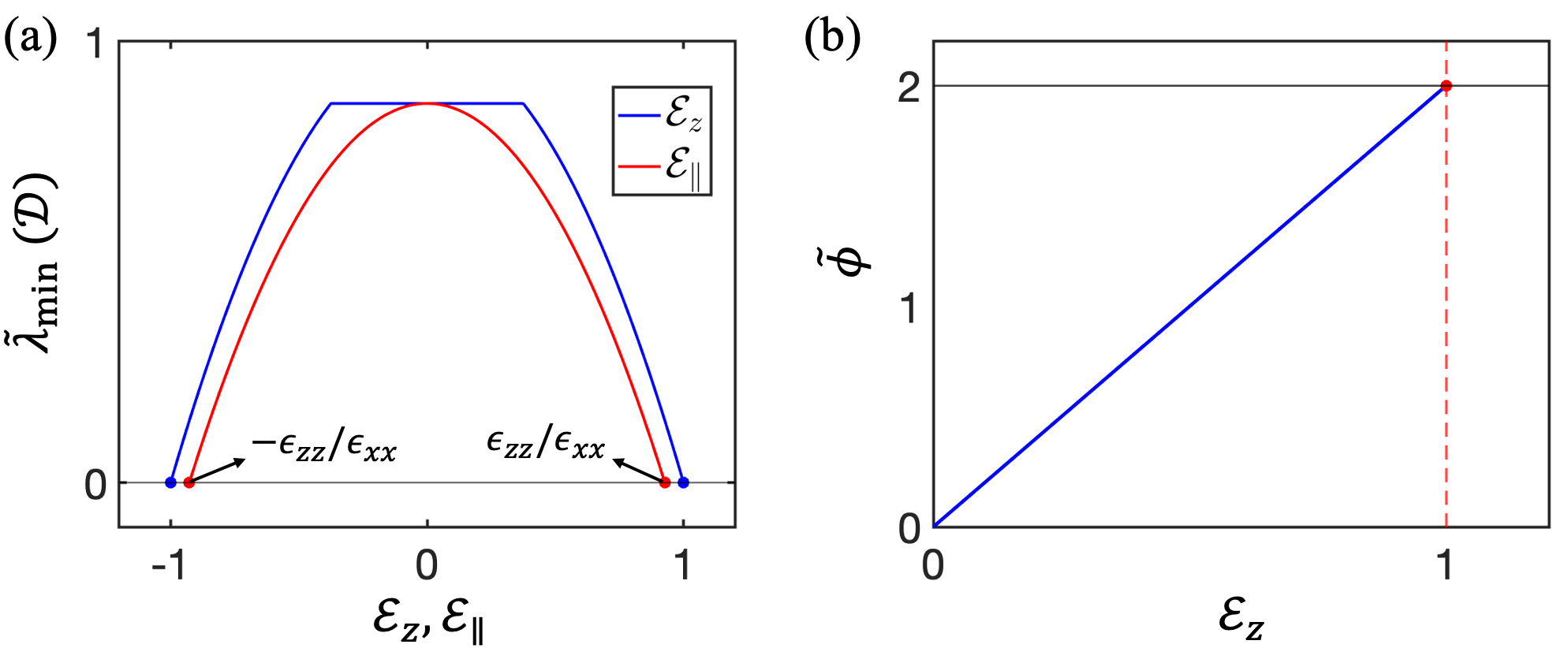}
\caption{(a) The dimensionless lowest eigenvalue $\widetilde{\lambda}_{\rm min}$ for the $d$-wave altermagnet as a function of $\mathcal{E}_z$ (blue, $\mathcal{E}_\parallel=0$) and $\mathcal{E}_\parallel$ (red, $\mathcal{E}_z=0$). (b) The rescaled phase retardance $\widetilde{\phi}$ as a function of $\mathcal{E}_z$ for $\mathcal{E}_\parallel=0$.}
\label{fig:RuO}
\end{figure}

\begin{figure*}
\includegraphics[width=\textwidth]{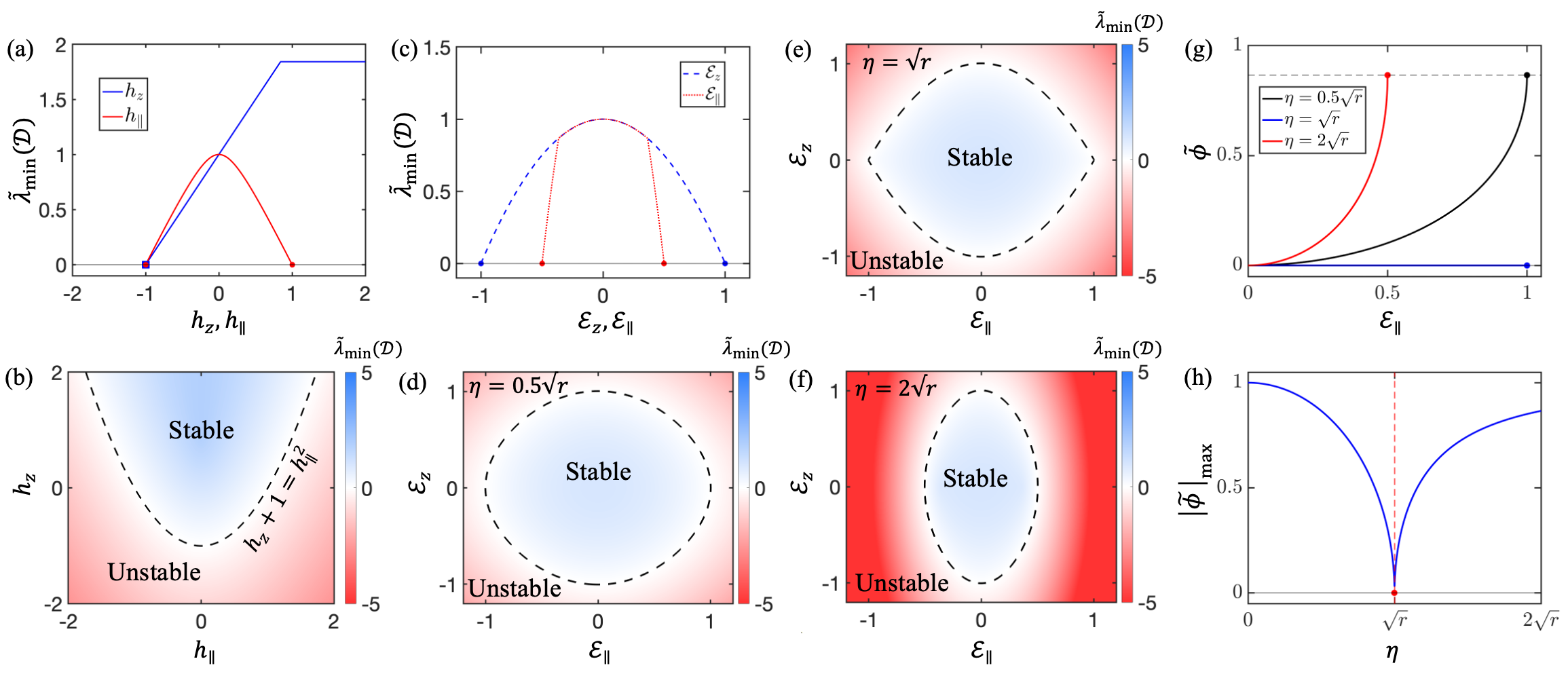}
\caption{(a) The dimensionless lowest eigenvalue $\widetilde{\lambda}_{\rm min}$ for the M-type hexaferrite as a function of $h_z$ (blue, $h_\parallel=0$) and $h_\parallel$ (red, $h_z=0$). (b) The magnetic-field stability phase diagram in the $h_z$--$h_\parallel$ plane, with stable (unstable) regions shown in blue (red); the dashed black line denotes the nonlinear Tellegen boundary. (c) The dimensionless lowest eigenvalue $\widetilde{\lambda}_{\rm min}$ as a function of $\mathcal{E}_z$ (blue, $\mathcal{E}_\parallel=0$) and $\mathcal{E}_\parallel$ (red, $\mathcal{E}_z=0$). (d)--(f) Electric-field stability phase diagrams in the $\mathcal{E}_z$--$\mathcal{E}_\parallel$ plane for (d) $\eta=0.5\sqrt{r}$, (e) $\eta=\sqrt{r}$, and (f) $\eta=2\sqrt{r}$; dashed black lines denote the nonlinear Tellegen boundaries. (g) The rescaled phase retardance $|\widetilde{\phi}|$ as a function of $\mathcal{E}_\parallel$ for $\mathcal{E}_z=0$ and $\eta=0.5\sqrt{r}$ (black), $\eta=\sqrt{r}$ (blue), and $\eta=2\sqrt{r}$ (red). (h) The maximum phase retardance $|\widetilde{\phi}|_{\rm max}$ as a function of the nonlinear magnetoelectric anisotropy $\eta$.}
\label{fig:Hexa}
\end{figure*}

(ii) \textit{Hexagonal ferrite}. --- M-type hexagonal 
ferrites exhibit invariance under the magnetic point group $6/mm^{\prime}m^{\prime}$, 
with generators $\hat{C}_{6z}$, 
$\hat{m}_x\hat{T}$, and $\hat{P}$. The nonzero nonlinear 
magnetoelectric coefficients are $\beta_{xxz} = \beta_{xzx} = 
\beta_{yyz} = \beta_{yzy} \equiv \beta_1$, 
$\beta_{zzz} \equiv \beta_2$, and 
$\beta_{zxx} = \beta_{zyy} \equiv \beta_3$. Since 
$\beta_3 \gg \beta_2$ \cite{popov2022plane}, we neglect $\beta_2$ henceforth. 
Unlike the $d$-wave altermagnetic case, the applied field $H_z^{(0)}$ 
renormalizes the diagonal elements of $\tilde{\epsilon}$ as 
$\tilde{\epsilon}_{xx} = \tilde{\epsilon}_{yy} = 
\epsilon_{xx} + \beta_3 H_z^{(0)}$ and 
$\tilde{\epsilon}_{zz} = \epsilon_{zz} + \beta_2 H_z^{(0)}$, 
without generating an off-diagonal magnetoelectric tensor 
$\alpha_{ij}$ (see Sec. S3 in SM for details \cite{SM}). For the M-type hexaferrite, we first consider a purely magnetic
background field, $\mathbf{E}^{(0)}=0$. We define
$h_\parallel^2=\beta_1^2[(H_x^{(0)})^2+(H_y^{(0)})^2]/
(\epsilon_{xx}\epsilon_{zz})$,
$h_z=\beta_3H_z^{(0)}/\epsilon_{xx}$, and
$r=\epsilon_{zz}/\epsilon_{xx}$. The rescaled eigenvalues
$\widetilde{\lambda}=\lambda/\epsilon_{xx}$ are
$\widetilde{\lambda}_1=1+h_z$ and
$\widetilde{\lambda}_{\pm}
=[1+h_z+r\pm\sqrt{(1+h_z-r)^2+4rh_\parallel^2}]/2$.
Fig.~\ref{fig:Hexa}(a) shows the lowest eigenvalue
$\widetilde{\lambda}_{\rm min}$ for fields applied separately along
the out-of-plane and in-plane directions. For $h_z=0$, increasing
$|h_\parallel|$ lowers $\widetilde{\lambda}_{\rm min}$ until the
nonlinear Tellegen limit is reached at $h_\parallel=\pm1$. In contrast,
for $h_\parallel=0$, a negative $h_z$ drives the system to the
Tellegen limit at $h_z=-1$, whereas a positive $h_z$ initially
increases the lowest eigenvalue and therefore stabilizes the medium.
For sufficiently large positive $h_z$,
$\widetilde{\lambda}_{\rm min}$ saturates at $r$. The competition
between the two field components is shown in
Fig.~\ref{fig:Hexa}(b). The stable and unstable regions
are separated by the dashed black line, $h_\parallel^2=1+h_z$, showing that a positive
out-of-plane field increases the in-plane field required to reach the
nonlinear Tellegen limit. We next consider a purely electric background field,
$\mathbf{H}^{(0)}=0$, and define the rescaled fields as
$\mathcal E_z=\beta_1E_z^{(0)}/\sqrt{\epsilon_{xx}\mu_{xx}}$,
$\mathcal E_\parallel=\beta_3E_\parallel^{(0)}/
\sqrt{\epsilon_{xx}\mu_{zz}}$ and the nonlinear magnetoelectric anisotropy as
$\eta=(\beta_1/\beta_3)\sqrt{\mu_{zz}/\mu_{xx}}$. Fig.~\ref{fig:Hexa}(c) shows the one-dimensional
electric-field dependence for $\eta=2\sqrt r$. For a purely
out-of-plane field, $\widetilde{\lambda}_{\rm min}
=\min(r,1-\mathcal E_z^2) = 1-\mathcal E_z^2$ since $r>1$ for BaFe$_{12}$O$_{19}$ \cite{shen2016quantum}, and the nonlinear Tellegen limit occurs
at $\mathcal E_z=\pm1$. For a purely in-plane field,
$\widetilde{\lambda}_{\rm min}
=\min(1-\mathcal E_\parallel^2,
r-\eta^2\mathcal E_\parallel^2)$, so that the critical field is
$|\mathcal E_\parallel^*|
=\min(1,\sqrt r/|\eta|)$. For the value
$\eta=2\sqrt r$ used in Fig.~\ref{fig:Hexa}(c), the in-plane
Tellegen limit is reached already at
$|\mathcal E_\parallel|=1/2$. For in-plane electric fields $|\mathcal{E}_\parallel| < \sqrt{(r-1)/(\eta^2-1)}$, $\widetilde{\lambda}_\text{min}$ has the same behavior as the out-of-plane case, as seen by the merging of the red and blue lines in Fig.\ref{fig:Hexa}(c). The full electric-field dependence is shown in
Figs.~\ref{fig:Hexa}(d)--\ref{fig:Hexa}(f) for
$\eta=0.5\sqrt r$, $\sqrt r$, and $2\sqrt r$, respectively.
In each panel the dashed contour denotes
$\widetilde{\lambda}_{\rm min}=0$. The stability conditions reduce to
$1-\mathcal E_z^2-\mathcal E_\parallel^2>0$ and
$r(1-\mathcal E_z^2-\mathcal E_\parallel^2)>
\eta^2\mathcal E_\parallel^2(1-\mathcal E_\parallel^2)$. $\eta$ determines the nonlinear Tellegen limit along the in-plane direction whereas the out-of-plane stability is unaffected by $\eta$. For $\eta<\sqrt r$ (Fig.~\ref{fig:Hexa}(d)), the stable region
extends to $|\mathcal E_\parallel|=1$. At
$\eta=\sqrt r$ (Fig.~\ref{fig:Hexa}(e)), the two electric-field
channels become equally competitive, while for
$\eta>\sqrt r$ (Fig.~\ref{fig:Hexa}(f)) the stable region becomes
compressed along the in-plane direction and the Tellegen limit is
reached at $|\mathcal E_\parallel|=\sqrt r/|\eta|<1$.
For $\eta=2\sqrt r$, this gives
$|\mathcal E_\parallel|=1/2$. Thus, both the
stability region in electric-field space and the most efficient
direction for approaching the nonlinear Tellegen limit are controlled
by the anisotropy of the nonlinear magnetoelectric coupling.

\textit{Electrically Tunable Linear Birefringence}. Linear birefringence, in which two orthogonal linearly polarized modes acquire different refractive indices, is a fundamental phenomenon in crystal optics \cite{born1964principles}. Here, the nonlinear magnetoelectric coupling provides a distinct mechanism for electrically tunable birefringence. For the $d$-wave altermagnet, an out-of-plane static electric field $E_z^{(0)}$ induces diagonal magnetoelectric components $\alpha_{xx}=\beta E_z^{(0)}$ and $\alpha_{yy}=-\beta E_z^{(0)}$. The opposite signs of these two components lift the degeneracy of the two orthogonal linear-polarization eigenmodes and generate a phase retardance that is linear in the applied electric field. For a probe propagating along $\hat z$, the full Maxwell derivation is given in Sec. S4.A of the SM \cite{SM}. Defining the rescaled phase retardance $\widetilde{\phi}=(c'/\omega)(\Delta\phi/L)$, with $c'=1/\sqrt{\epsilon_{xx}\mu_{xx}}$, we obtain $\widetilde{\phi}=2\mathcal E_z$. As shown in Fig.~\ref{fig:RuO}(b), the retardance therefore increases linearly with $|\mathcal E_z|$ and reaches the limiting value $|\widetilde{\phi}|=2$ at the nonlinear Tellegen boundary $|\mathcal E_z|=1$. The optical response of the M-type hexaferrite is qualitatively different. An out-of-plane electric field does not produce linear birefringence for propagation along $\hat z$, whereas an in-plane field $E_\parallel^{(0)}$ does. In this case the birefringence is controlled not only by the applied field but, strikingly, by the ratio $\eta=(\beta_1/\beta_3)\sqrt{\mu_{zz}/\mu_{xx}}$ of the nonlinear magnetoelectric couplings. The exact mode dispersions are derived in Sec.~S4.B of the SM \cite{SM}. The rescaled phase retardance is $\widetilde{\phi}=\sqrt{1-\mathcal E_{\parallel}^2}-\sqrt{1-(\eta^2/r)\mathcal E_{\parallel}^2}$. Fig.~\ref{fig:Hexa}(g) shows $|\widetilde{\phi}|$ for $\eta=0.5\sqrt r$ (black), $\sqrt r$ (blue), and $2\sqrt r$ (red). At $\eta=\sqrt r$, the two modes remain degenerate and the birefringence vanishes identically; away from this point, a finite electrically tunable retardance develops. Stability up to the nonlinear Tellegen limit requires $|\mathcal E_{\parallel}|<\min(1,\sqrt r/|\eta|)$, which directly bounds the optical response. From Fig.~\ref{fig:Hexa}(g), we observe that $\eta=0.5\sqrt r$ and $\eta=2\sqrt r$ share the same upper bound. Thus, we plot the optical upper bound $|\widetilde{\phi}|_\text{max}$  a a function of the nonlinear magnetoelectric anisotropy $\eta$ in Fig.~\ref{fig:Hexa}(h). The bound vanishes at $\eta=\sqrt{r}$, where the birefringence is identically zero. Away from this point, the upper bound exhibits the duality $|\widetilde{\phi}_{\max}(\eta/\sqrt{r})|=|\widetilde{\phi}_{\max}(\sqrt{r}/\eta)|$, with $\eta=\sqrt{r}$ as the self-dual point.

\textit{Discussion}.
We have introduced and characterized the nonlinear Tellegen 
limit — a field-tunable electromagnetic instability arising 
from cubic magnetoelectric coupling in centrosymmetric 
magnetically ordered materials. The stability of a nonlinear 
magnetoelectric medium is governed by the positive definiteness of $\mathcal{D}$, or equivalently $\lambda_{\rm min}(\mathcal{D})>0$,
where the induced magnetoelectric tensor is entirely controlled 
by an applied static field, providing a direct, continuously tunable handle on the electromagnetic stability 
of the medium — a capability absent in the linear 
magnetoelectric regime. Two material examples illustrate the breadth of this 
phenomenon. In $d$-wave altermagnets like MnF$_2$, the symmetry-allowed $\beta$ tensor drives the system toward the nonlinear 
Tellegen limit at two different electric field strengths depending on the orientation of the applied electric fields. The resulting linear birefringence is electrically tunable and reaches the universal bound $|\widetilde{\phi}|_{\rm max}=2$ at the nonlinear Tellegen limit. For $f\sim1$ THz and a film thickness $L\sim100$ nm, this corresponds to a phase retardance of order $10^{-2}$ rad for MnF$_2$, well within the sensitivity of THz polarization-resolved spectroscopy~\cite{morris2012polarization,katletz2012polarization}. For MnF$_2$, the critical electric fields are $E_z^*=\sqrt{\epsilon_{xx}\mu_{xx}}/|\beta|$ and $E_\parallel^*=\sqrt{\epsilon_{zz}\mu_{xx}}/|\beta|$, reflecting the directional dependence of the nonlinear Tellegen limit. Using $\epsilon_{xx}/\epsilon_0\simeq8.9$, $\epsilon_{zz}/\epsilon_0\simeq7.65$ \cite{seehra1986effect}, and $\mu_{xx}\simeq\mu_0$, these correspond to $E_z^*\simeq9.9\,({\rm ns/m})/|\beta|$ and $E_\parallel^*\simeq9.2\,({\rm ns/m})/|\beta|$. To our knowledge, the static cubic magnetoelectric coefficient $\beta$ relevant here has not been calculated for MnF$_2$. A first-principles determination of this coefficient would provide a fully quantitative prediction for the critical electric fields and represents a natural next step. In hexagonal ferrites, the richer $\beta$ tensor structure produces an even broader field-tunable stability landscape. Magnetic fields exhibit a striking asymmetry between stabilizing and destabilizing directions, while electric fields can drive the system to the nonlinear Tellegen limit along both in-plane and out-of-plane directions. Using $\epsilon_{xx}\simeq19\epsilon_0$ \cite{shen2016quantum} and the nonlinear coefficient $\beta_3$ inferred from the current-driven FMR measurements of Ref.~\cite{popov2022plane}, the corresponding effective field scale is of order $10^{-1}$--$10$ Oe, depending strongly on sample type. The inferred Tellegen field scale is far below the intrinsic magnetic scales of BaFe$_{12}$O$_{19}$: the magnetocrystalline anisotropy field is $H_a\sim17$ kOe, the demagnetizing scale is of order $4$--$5$ kOe, and the corresponding uniform FMR frequency is approximately $46$--$47$ GHz \cite{popov2022plane}. The nonlinear coefficient of Ref.~\cite{popov2022plane} was moreover extracted under large magnetic bias fields, $8$--$9.5$ kOe. Thus coherent magnetization reorientation or softening of the uniform FMR mode is unlikely to occur at an Oe-scale field. Domain and coercive-field scales, however, are strongly sample dependent: the coercivity is nearly negligible in high-quality single crystals, while polycrystalline samples can range from a few Oe to several kOe depending on grain-boundary pinning \cite{dho2005effects}. We therefore cannot exclude domain reconfiguration from preempting the electromagnetic instability in a given sample, and regard the estimated $H_z^*$ only as an effective order-of-magnitude scale. We also note that $\beta_3$ is inferred from current-driven FMR measurements \cite{popov2022plane} rather than from a direct equilibrium measurement of the static cubic free-energy coefficient. Its mapping to the thermodynamic $\beta_3$ used here may therefore involve additional dynamical and sample-dependent effects. The coefficient $\beta_1$, characterizing the $H_xE_xE_z$ coupling, has not yet been experimentally determined. In addition to setting the in-plane magnetic Tellegen limit $H_\parallel^*=\sqrt{\epsilon_{xx}\epsilon_{zz}}/|\beta_1|$, the ratio $\beta_1/\beta_3$ controls the electric-field stability landscape and the maximum linear birefringence. A direct measurement or first-principles calculation of $\beta_1$ would therefore provide a quantitative test of several predictions of the present theory.

The present framework applies to any centrosymmetric 
magnetically ordered material with a nonzero $\beta_{ijk}$ 
tensor, encompassing a wide class of altermagnets, 
hexaferrites, and multiferroics. Among altermagnets, 
$\alpha$-MnTe
also forbids linear magnetoelectric coupling 
and can permit the cubic coupling $\beta_{ijk}$, and 
has been unambiguously confirmed as altermagnetic~\cite{osumi2024mnte,lee2024broken,
amin2024nanoscale,hariki2024xmcd}. In metamaterials, $\epsilon$ and $\mu$ can 
be engineered independently~\cite{veselago1967electrodynamics,smith2004metamaterials,
soukoulis2011metamaterials}, making it possible in principle to engineer substantially reduced critical fields $E^*$ and $H^*$, thereby bringing the nonlinear Tellegen instability into experimentally accessible regimes. The nonlinear Tellegen limit thus establishes a new 
organizing principle at the intersection of magnetic 
space group symmetry, nonlinear electromagnetic response, 
and fundamental stability bounds — with concrete 
experimental signatures accessible to current 
measurements and a rich landscape of open questions for 
future exploration.

\textit{Acknowledgments} --- The authors thank Saswata Mandal for helpful discussions. The authors acknowledge the use of GPT-5 (OpenAI), directed through author-written prompts, for assistance with manuscript editing, code organization, and analytical derivation checks, and Claude 4 (Anthropic) for assistance in preparing Figure 1; all AI-assisted code, calculations, derivations, figures, scientific conclusions, and final manuscript content were independently checked by the authors and remain the responsibility of the authors.


%

\end{document}